# Ontology based Approach for Semantic Service Selection in Business Process Re-Engineering

S. CHHUN[1], N. MOALLA[1], Y. OUZROUT[1]
1  {sophea.chhun, nejib.moalla,yacine.ouzrout}@univ-lyon2.fr

**Abstract.** This research aims to provide the possibility to the business analysts to be able to know whether their design business processes are feasible or not. In order to solve this problem, we proposed a model called BPMNSemAuto that makes use of the existing services stored in the service registry UDDI (Universal Description Discovery and Integration). From the data extracted from the UDDI , the WSDL files and the tracking data of service execution on the server, a Web Service Ontology (WSOnto) is generated to store all the existing services. The BPMNSemAuto model takes an input of business process design specifications, and it generates an executable business process as an output. It provides an interface for business analysts to specify the description of each service task of the design business process. For each service task, the business analysts specify the task objective (keywords), inputs, outputs and weights of the QoS (Quality of Service) properties. From the design business process with the service task specifications, a Business Process Ontology (BPOnto) is generated. A service selection algorithm performs the mapping between the instances of the WSOnto and the BPOnto to obtain possible mappings between these two ontologies. The obtained mappings help the model to acquire web services to execute the desired service tasks. Moreover, the consistency checking of the inputs of the proposed model is performed before executing the service selection algorithm. WordNet is used to solve the synonym problems and at the same time a keyword extraction method is presented in this paper.

## 1.1 Introduction

The Workflow Management Coalition provides a definition of Business Process, saying that "it is a set of one or more linked procedures or activities which collectively realize a business objective or policy goal, normally within the context of an organizational structure defining functional roles and relationships" [1]. Rummler and Brache defined business process as "the series of steps that a business executes to produce a product or service" [2]. The business process applications can be modeled with different modelling specifications such as BPMN (Business Process Model and Notation) [3], Petri Net [4,5], Workflow and UML (Unified Modeling Language). This research study focuses on the modelling of business processes with BPMN specifications (BPMN2.0). Some examples of business process modeled with BPMN are presented in [6], such as "handling and invoicing process application", "taxi reservation application" and "online purchasing application". Correia and Abreu [7] state that "A BPMN2 process model diagram has around 100 different modeling constructs, including 51 event types, 8 gateway types, 7 data types, 4 types of activities, 6 activity markers, 7 task types, 4 flow types, pools, lanes, etc." However, this research work considers only the automatic implementation of the service task (one kind of task type) by using the existing services stored in the service registry UDDI (Universal Description Discovery and Integration).

A service task can be performed by a web service or a composite service. The web service is a software module created to perform a specific business task. It is described by the service description languages such as WSDL (Web Service Description Language) and WSDL-S (Web Service Description Language-Semantic). These languages provide different capabilities, for example, WSDL cannot store the pre-condition and post-condition of a service but WSDL-S and OWL-S do.

Researchers use the ontology to represent the semantic meaning of services and as a knowledge base. The Ontology is an explicit specification and hierarchy of different concepts. It defines properties, characteristics and behaviors of objects in the same domain; and it expresses the relationships between concepts [8]. An ontology consists of three elements: vocabularies, specifications and constraints. The vocabularies describe the domain of ontology and the constraints are used to capture knowledge about the ontology's domain. The specifications define the relationships between different concepts of the ontology. Moreover, some ontology languages have been proposed such as RDF, DAML-OIL, WSMO, OWL and OWL-2. This research study builds a Web Service Ontology for the semantic representation of the services stored in the service registry (UDDI).

Globally, this research study aims to provide an automatic implementation of business processes by re-using the existing web services stored in the service registry. A model called BPMNSemAuto is proposed and takes the input of business process design specifications by the users; and it generates an output of an executable business process. After designing the business processes, the business analyst provides the specifications of each service task through a user interface such as context, inputs, outputs and weight. After that the BPMNSemAuto model performs the service selection and composition to choose the most suitable services to execute every service task. The service selection is done by comparison between the ontology represents the user's requirements and Web Service Ontology; and it uses the QoS (Quality of Service) values to rank the matched services.

This work targets some research problems such as: (i) Semantic representation of the existing web services and the users' requirements. It is because the service registry UDDI (Universal Description, Discovery and Integration) supports only keywords matching and does not store the non-functional properties of web services. However, the non-functional properties of services are the important criteria of the service selection algorithm. (ii) A service selection and composition algorithm. (iii) A solution to solve the problems of synonyms because organizations usually use their own specific terms to name business elements and web services.

The rest of this paper is organized as follows. Section 2, "Related Works", presents the current existing solutions related to this research study. Section 3, "Proposed Solution", introduces the proposed model architecture, the web service ontology structure and a keywords extraction method. Finally, this paper is finished by a conclusion and future work.

## 1.2 Related Works

### 1.2.1. Business Process Modeling and Modeling Languages

Business Process Modelling Notation (BPMN) is a standard notation for modeling the business processes. It bridges the gap between the design and the implementation of the business processes. The primary goal of BPMN is "to provide a notation that is readily understandable by all business users, from the business analysts that create the initial drafts of the processes, to the technical developers responsible for implementing the technology that will perform those processes, and finally, to the business people who will manage and monitor those processes"[3]. BPMN represents the business process as a Business Process Diagram (BPD)[9]. It divides the elements of business process into four categories: (i) Flow objects that define the behaviors of a business process. A flow object can be an event, activity and gateway; (ii) Connecting objects that connect between two flow objects or between a flow object with other resources. Three types of connecting objects exist: sequence flow, message flow and association; (iii) Swimlanes that group the primary modeling elements. There are two kinds of swimlanes, pool and lane; (iv) Artefacts allow to provide additional information about the process. Artefacts are sub-categorized into data object, group and annotation. In addition, BPMN permits the automatic translation of the graphical business process into BPEL (Business Process Execution Language).

sBPMN (Semantic Business Process Modelling Notation) ontology provides the semantic meaning of each element of the business processes, allows machine readable, and allows reasoning on the process description [10]. In [9], the authors concluded that a modeling language is chosen based on some criteria such as modeling approaches ( graph based or rule based) and capabilities of the language (expressibility, flexibility, adaptability, dynamism and complexity).

In summary, the graphical process modeling approach is more used than the rule based approach because it provides a graphical interface that allows business users to be able to model their business processes.

### 1.2.2. Ontology Representation

In the literature reviews, experts generally define the hierarchy of ontologies and design ontologies for a specific domain of applications. This manual hierarchy of ontology structure is supposed to provide a better accurate and comprehensive representation of the domain information; because experts understand the domain of applications very well. The ontology building for a specific domain is easier than building a generic one that shares amount many application domains. The advantage of the generic domain ontology is the independent aspect of applications. It is rich in axioms, but a heavyweight ontology. The generic ontologies must complete some constraints such as modelling expressiveness, clear semantics and rich formalization, adaptability, harmonization with other standards and usability [11].

In addition, ontology is used to improve the semantic representation of web services in the service oriented architecture. It supports the service selection and composition process, and provides the ability to determine different matching degree between two concepts such as exact, plugin, subsume, intersection and disjoint [12].

Different ontology languages are proposed and they must verify a number of requirements in order to be useful for the business system modelling such as: well defined semantics, sufficient expressive power, powerful and understandable reasoning mechanisms, easy to use with reasonable compact syntax [13]. Each ontology language is different from each others by their expressiveness, supported data type of concept's properties, syntax, constraint checking, top level elements and its ability to support reasoning. Azleny et al [14], state that an ontology language is selected based on four criteria: intended use, expressiveness, automated reasoning and user perception. In addition, the continuous evolution of ontology languages is also a main criterion for choosing an ontology language.

### 1.2.3. Quality of Service (QoS)

In order to improve the accuracy of the result of the service selection and composition algorithms, the non functional properties of services has to be considered and not only their functional properties. The attributes of QoS are mainly defined for a specific domain and are categorized into groups based on their characteristic such as performance, security and context [15][16]. It is hard for the users to define the value of each attribute of QoS, therefore it is easier for them if they just provide the weight value. The weight defines the importance level of each attribute of QoS [17].

However, it requires addition work to do when working with the QoS because the limitations of the current web service technologies. For example, the representation of the services with WSDL and OWL-S does not allow to express the QoS values. The service registry UDDI does not support the storing of QoS values. There is no standard structure of the QoS ontology and how to calculate them.

## 1.3 Proposed Solution

This research aims to provide a solution for the automatic implementation of the business processes from business process design specifications by reusing the existing services stored in the service registry. A model called BPMNSemAuto is proposed in order to solve this problem. The input of BPMNSemAuto model is the business process design specifications (bpmn file format) that can be designed with any supporting editor of BPMN specifications (ex. Jdeveloper and Eclipse). The output of BPMNSemAuto model is the executable business process corresponding to the business process designed by the users. In addition, the BPMNSemAuto model performs only the automatic implementation of the service task and not the other business tasks specified by the users.

### 1.3. 1  BPMNSemAuto Model Architecture

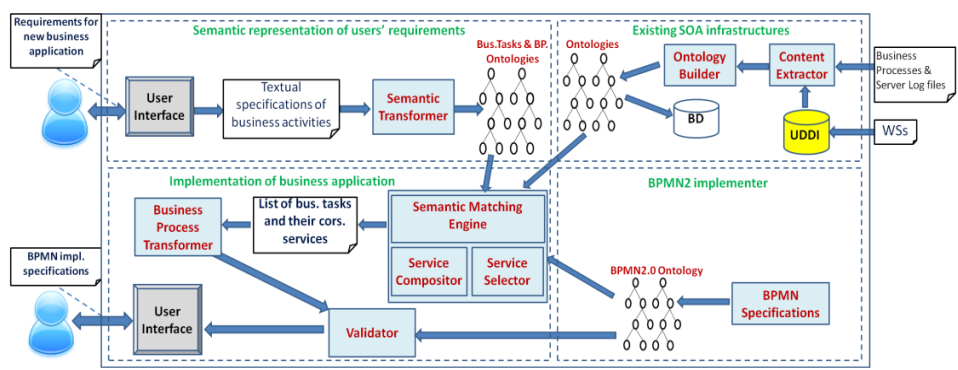

Fig 1. 1. BPMNSemAuto model architecture

The proposed model architecture called BPMNSemAuto (see Fig 1.1) is described in detail step by step as follows:
- First, the uses design the business process with any supporting editors of BPMN specifications, and it is used as the input for BPMNSemAuto model.
- After that the BPMNSemAuto model provides an interface to the users for specifying the description of each service task including context, inputs, outputs and weights to identify the importance of the QoS properties.
- The BPMNSemAuto checks the data type consistency of the input and output specifications from the users. If any inconsistency is detected, it generates an alert to the users.
- The BPMNSemAuto model builds two different ontologies, a BPOnto ontology to represent the specifications of business process designed by the users; and a WSOnto ontology to represent all the existing services stored in the service registry UDDI.
- Then, the Semantic Matching Engine performs the instance matching between BPOnto and WSOnto ontology to obtain the possible matched services to perform the requested service tasks. It ranks the matched services based on the QoS values. Moreover, it performs the service composition algorithm to create the composite services if the existing atomic services cannot reply to the requirements of users.

- After that, the Business Process Transformer generates an executable business process corresponding to the designed business process of the users.
- Next, the Validator validates the generated business process to check the syntax inconsistency with the support of the BPMN 2.0 ontology defined in (18). It corrects the syntax error if possible, if not it alerts to the users.

**1.3. 2  Web Service Ontology**

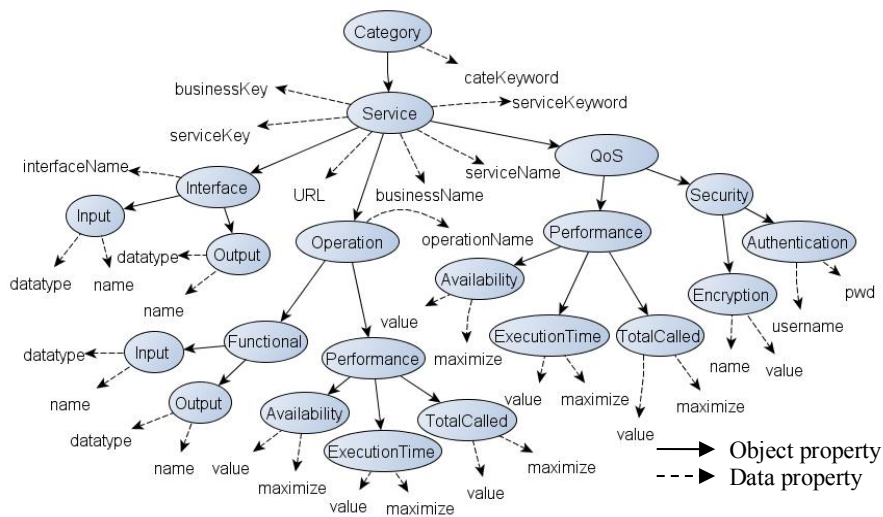

Fig 1. 2. Web Service Ontology

The Web Service Ontology (WSOnto) is proposed to store all the existing services grouped in categories. This WSOnto provides enough information about the services for the service selection algorithm and for the implementation of services in the business process generation process. The service category is defined by the values of tModel of the UDDI. The Machine Learning Techniques such as Support Vector Machines (SVM) [19] and Nearest Semantic Similarity [20] require the training data in order to define the category patterns. However in this research study, the existing services are already published in the specific categories by publishers. In the future, the machine learning techniques might be needed for suggesting the category of service to the service publishers when they publish their services. Each service of the WSOnto is described by its functional and non-functional properties. The functional properties of services are specified by their service interface (for calling the service) and operations. The non-functional properties of services are specified by the QoS and some additional information such as service-key, business-entity-key and WSDL file location. The complete information about services is useful when inquiring, matching and ranking the services. The QoS values are used to rank the matched services because they can improve the re-usability of services and reduce the development cost. The content of the WSOnto ontology is extracted from the UDDI registry (serviceKey, businessKey, WSDL file location, businessName and service's security information), WSDL files (operations and interface of the services) and the tracking data of service execution on the server (performance value of the service and service's operations). The WSOnto ontology (see Fig 1.2) contains of twenty one classes, twenty object properties and thirty three datatype properties.

### 1.3. 3  Text Extraction

In the proposed WSOnto ontology (Fig 1.2), each service and service category are linked to a list of keywords. The keywords of category are extracted from the description of tModel, description of service and name of service of all services in the same category. The keywords of service are extracted from the information of service (description, name, name of the service's operation, description of the service's operation, name of the inputs and name of the outputs).

By adopting the method presented in [19] and [20], the keywords are extracted by the following steps (Fig 1.3):
1. **Extract Keyword**: From the text description, POS (Part Of Speech) tagger and Tokenizer are used to extract the words that have part of speech as noun, compound noun and verb.
2. **Split combined terms**: From the keywords generated in step 1, the keyword extraction module splits the keyword if the keyword is a combined word.
3. **Remove stop words**: This sub module removes the stop words, those stop words can be an article, proposition and some useless words such as service, operation, wsdl and soap.
4. **Word stemming**: It is a process to find the originality of a word. In English language, the nouns and verbs can be in singular and plural form, but they mean the same thing. Therefore, the best solution is just to store the infinitive form of the word.

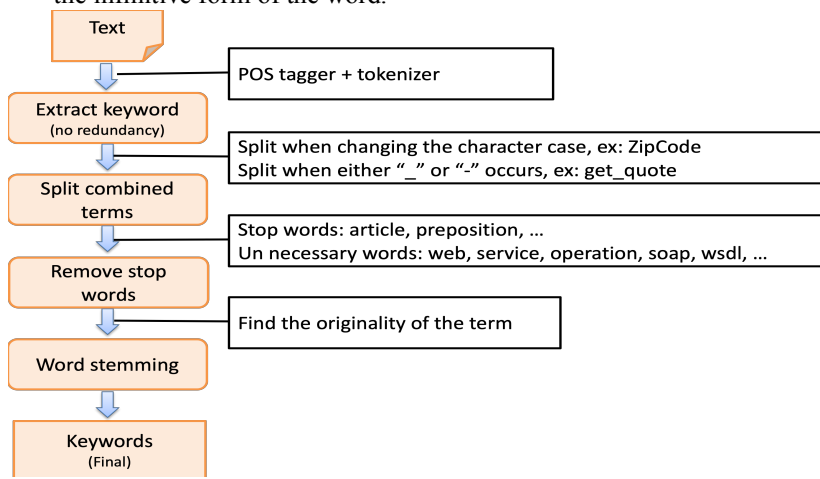

Fig 1. 3. Keyword extraction method

### 1.3.4 WordNet

WordNet is an English online lexical reference system which provides synonym, hypernyms (generalization) and hyponyms (specialization) sets consisting of nouns, verbs, adjectives and adverbs. At the same time, it provides APIs (Application Programming Interface) in different programming languages that allow us to query to the server to obtain a list of synonym words. The proposed BPMNSemAuto model uses WordNet to obtain the synonym terms. Using WordNet can solve the synonym problems cause by the use of specific terms to name the concepts specified by the companies.

### 1.3.5 Consistency Checking

Through the provided interface, the users are responsible for designing the business process and specifying the description of each service task based on a specific format provided. They have to provide the objective of the service tasks (in the form of keywords), inputs, outputs and the weights of the QoS attributes. Before performing the service selection algorithm, the BPMNSemAuto model checks the consistency of the design business process by comparing between the data type of the outputs of a service task with the data type of inputs of another service task in the sequence. The data type can be a simple type (integer, float, string) or a complex data type (object data type). The object data type is the composition of many simple data types. Therefore, the checking of data type consistency is the comparison of simple data types. For example, if an output is a string but the input is a float, then it shows an inconsistency. However, if the output is an integer and the input is float, then it is acceptable. This consistency checking process is really important, if it is not considered, maybe the generated business process cannot be executed.

### 1.3.6 Service Selection

From the user specifications of each service task, the correspondent Service Task Ontology (STOnto) is generated, or a Business Process Ontology (BPOnto) is generated to represent the design business process of the user. After that the BPMNSemAuto model makes the comparison between the STOnto or BPOnto and the WSOnto ontology to obtain possible mappings between these two ontologies. The obtained mappings help the model to acquire web service to execute the desired service task.

The ontology matching methods focus on two things, schema matching and instance matching. A survey of ontology matching tools and techniques is presented in [21]. However in this research study, the instance matching is focused. The OWL API is used to traverse the two ontologies in order to compare the corresponding individuals. The service selection algorithm matches first the keywords, then the inputs and the outputs (name and data type) with the support from WordNet dictionary. After obtaining the list of matched services, an existing service ranking algorithm presented in [22] is adopted to rank them. The authors of [22] presented a Multi-dimensional Multi choice 0-1 Knapsack Problem (MMKP) to choose the best solution out of the K groups of items. However, the algorithm is reduced to just apply with only one group of services that provide the same functionality. Therefore, the value of the utility function of each service can be calculated with the equation 1.1.

$$F = \sum_{i=1}^{\alpha} w_i * \left(\frac{q_{ai}-\mu_{ai}}{\sigma_{ai}}\right) + \sum_{j=1}^{\beta} w_j * \left(1 - \frac{q_{bj}(K)-\mu_{bj}}{\sigma_{bj}}\right) \quad (1.1)$$

Where α: the number of QoS properties that are required to maximize their values; β: the number of QoS properties that are required to minimize their values; w: weight of each QoS parameter that is set by users (0< $w_i$, $w_j$ <1); μ and ϭ are the average value and the standard deviation of QoS attributes for all candidates in a service class; $\sum_{i=1}^{\alpha} w_i + \sum_{j=1}^{\beta} w_j = 1$; α+β = total number of QoS attributes; q: QoS value.

Finally, the service with the maximum value of the utility function F is selected. From the original business process design specifications with a list of services corresponding to the service tasks, an executable business process is generated. The BPMN2.0 ontology [18] is planned to use to validate the generated business process.

## 1.4 Conclusion & Future Work

A BPMNSemAuto model is defined to perform an automatic implementation of business processes from their design specifications and the existing services. A WSOnto Ontology is proposed to store all the necessary information of the existing services stored in the service registry in different categories. Moreover in order to check the usability of the WSOnto, the OWL API is used to traverse the WSOnto and STOnto or BPOnto ontology in order to compare the corresponding individuals to obtain the matched services; and an existing service ranking algorithm is adopted to rank the matched services. From different keyword extraction literature reviews, a synthesis of a keyword extraction method is presented.

Currently, the data description of the services is extracted from the WSDL files and UDDI data; therefore, it does not contain the value of the pre-condition and the post-condition. In the future work, OWL-S might be used to represent the services in order to have the value of the pre-condition and the post-condition.

**Acknowledgement**

This project has been funded with support from the European Commission (EMA2-2010- 2359 Project). This publication reflects the views only of the authors, and the Commission cannot be held responsible for any use which may be made of the information contained therein.